\documentclass[conference]{IEEEtran}
\usepackage[compatibility=false]{caption}

\usepackage{amsmath,amssymb,amsfonts}
\usepackage{graphicx}
\usepackage{enumitem}
\usepackage{textcomp}
\usepackage{booktabs}
\usepackage{tabularx}
\usepackage{url}
\usepackage{cite} 
\usepackage{array} 
\usepackage{booktabs}
\usepackage{float}
\usepackage{multirow}
\usepackage{array} 
\usepackage[para,flushleft]{threeparttable} 
\usepackage{tikz}
\usetikzlibrary{shapes, arrows, positioning, shapes.misc, calc, shapes.geometric, arrows.meta, fit, backgrounds, shadows}
\usepackage{subcaption}
\usepackage[labelfont=bf,compatibility=false]{caption}
\usepackage[font=footnotesize]{caption}

\begin{document}

\title{Forensic Text Recovery from Document Thumbnails Using Text-Aware U-Net}


\author{
    \IEEEauthorblockN{Mohit Ajaykumar Dhabuwala*}
    \IEEEauthorblockA{\textit{Cyber Diligence, Inc.} \\
    New York, USA \\
    md@cyberdiligence.com}
    
    \and
    \IEEEauthorblockN{Weifeng Xu}
    \IEEEauthorblockA{\textit{School of Criminal Justice} \\
    \textit{University of Baltimore}\\
    Maryland, USA \\
    wxu@ubalt.edu}
}

\maketitle

\begin{abstract}
Forensic analysis of low-resolution Word thumbnails presents a significant challenge because these artifacts may constitute the only remaining visual evidence of deleted documents. Standard Optical Character Recognition (OCR) tools perform poorly on such severely degraded images, creating a critical bottleneck in forensic evidence recovery. Although deep learning architectures, such as U-Net, can improve visual image quality, models optimized solely using pixel-based loss functions tend to favor visual smoothness over the structural fidelity required for accurate text recognition. To address this limitation, we propose a text-aware restoration framework designed for forensic triage. The proposed approach employs a U-Net architecture guided by a hybrid optimization strategy that integrates pixel-level fidelity with machine-readable text constraints. Specifically, an OCR-based Normalized Levenshtein Distance (NLD) metric is used during model selection to prioritize the recovery of an accurate character topology. Experimental results demonstrate that the proposed method significantly reduces the NLD, enabling reliable recovery of machine-readable text from severely degraded thumbnails.
\end{abstract}


\section{Introduction}
\label{sec:introduction}

In digital forensics, recovering legible text from fragmented artifacts is often pivotal. Microsoft Word documents frequently leave behind small bitmap ``thumbnail'' previews that persist even after the source file is deleted. As these low-resolution, compressed previews may represent the sole remaining visual evidence \cite{b20}, investigators rely on them to reconstruct timelines and corroborate digital traces.

However, these artifacts are not designed for readability. Aggressive downsampling and compression artifacts severely degrade textual details, confounding both human examiners and Optical Character Recognition (OCR) engines. Although deep learning frameworks such as U-Net \cite{b24} perform well in general image restoration \cite{b10,b21}, pixel-driven optimization (e.g., L1 loss) often favors visual smoothness over semantic precision. In a forensic context, this creates an unacceptable risk of ``hallucinating'' characters, generating factually incorrect text \cite{b3}.

We address this limitation by framing thumbnail reconstruction as a machine-readability task, rather than a purely aesthetic enhancement. Specifically, we aim to transform low-resolution, compression-degraded inputs into high-fidelity reconstructions that are verifiable using standard OCR tools. Although prior text-aware networks exist \cite{b15, b22}, many rely on computationally intensive sequential architectures that are unsuitable for rapid forensic triage. There remains a critical gap in the development of efficient convolutional models that directly embed legibility constraints into the training process.

This study proposes a Text-Aware U-Net tailored for forensic document enhancement, introducing a hybrid loss function that integrates pixel-level fidelity with an OCR-based objective. Unlike traditional methods that treat restoration and recognition as separate tasks, our approach forces the network to minimize character-level edit distances during restoration. This guides the model to generate outputs that are structurally optimized for automated textual analysis, addressing the following research questions:
\begin{itemize}
    \item \textbf{RQ1}: To what extent does the proposed text-aware model improve the visual quality (PSNR, SSIM) of severely degraded forensic thumbnails compared to standard baselines?
    \item \textbf{RQ2}: Does the integration of an OCR-based objective significantly improve text recoverability, as measured by Normalized Levenshtein Distance (NLD)?
\end{itemize}

The primary contributions of this study are as follows:
\begin{itemize}
    \item \textbf{Problem Formulation:} We formalize thumbnail reconstruction as a forensic task focused on OCR accuracy rather than purely visual aesthetics.
    \item \textbf{Hybrid Loss Function:} A novel objective combining L1 pixel loss with a penalty derived from OCR-based Levenshtein distance to enforce semantic integrity.
    \item \textbf{Empirical Validation:} We demonstrate that this paradigm significantly reduces error rates, establishing a robust method for recovering machine-readable evidence from low-resolution inputs.
\end{itemize}

The remainder of this paper is organized as follows: Section~II reviews the Related Work. Section~III details the Thumbnail Dataset. Section~IV presents the methodology. Section~V presents the Empirical Study and Results. Finally, Section~VI concludes with the key findings and directions for future research.

\section{Related Work}
\label{sec:related_work}

Restoring degraded Word Thumbnails remains a challenging problem in digital forensics, as it requires balancing visual enhancement with the semantic recoverability of textual content. This section reviews prior work on generic image restoration, adversarial super-resolution, text-aware enhancement, and OCR-guided optimization to contextualize the proposed approach.

Early advances in image restoration have established the effectiveness of convolutional neural networks for end-to-end super-resolution. Deep convolutional architectures have been shown to significantly outperform traditional interpolation techniques for image enhancement \cite{b1}. Building on this foundation, residual learning strategies were introduced to improve reconstruction accuracy by optimizing deep network structures without batch normalization \cite{b9}. Although these pixel-driven approaches achieve high reconstruction fidelity, they remain largely content-agnostic when applied to document images.

Adversarial learning models, such as generative adversarial networks (GAN), have demonstrated improved visual sharpness and perceptual realism through adversarial loss optimization \cite{b7}. Subsequent refinements of adversarial objectives produce visually convincing textures and edges \cite{b18}. However, GAN-based methods may introduce hallucinated structures because of their emphasis on perceptual realism, limiting their suitability for forensic document restoration, where factual textual accuracy is critical. Conditional adversarial models for image-to-image translation further illustrate this trade-off between visual realism and structural reliability \cite{b4}.

To address text-specific degradation, researchers have incorporated semantic constraints into restoration pipelines. It was shown that convolutional networks trained for text deblurring should be evaluated using OCR accuracy rather than pixel error alone, emphasizing machine readability as the primary objective \cite{b3}. OCR-driven super-resolution frameworks further integrate recognition feedback to guide enhancement iteratively \cite{b22}. Content-aware approaches that align visual reconstruction with recognition objectives have demonstrated improved preservation of character topology and text recoverability compared to generic enhancement methods \cite{b17, b2}.

The robustness of blind image restoration has also been explored. Gradient reliability mechanisms optimize convergence stability in blind restoration \cite{b23}. While residual dense U-Net variants provide a foundation for noise-resilient restoration \cite{b24}, they omit explicit text legibility constraints. Similarly, fast back-projection networks achieve noise-resistant imaging performance \cite{b14}.

In the forensic domain, reliable extraction of textual evidence from degraded thumbnails is essential for investigative validation \cite{b20, b6}. This study addresses this requirement by employing a text-aware U-Net architecture \cite{b12} guided by a metric-driven optimization strategy. Unlike GAN-based or recognition-dependent pipelines, the proposed fully convolutional framework prioritizes structural fidelity and text recoverability, thereby enabling efficient and verifiable forensic triage.

\section{Thumbnail Dataset}
\label{sec:dataset}

This study utilizes a curated subset of 1,000 document-thumbnail pairs derived from the synthetic dataset developed by E. Xu \textit{et al.} \cite{b20}. Samples were selected at a base resolution of $256 \times 256$ pixels and rendered in the Aptos typeface to establish a controlled evaluation environment\footnote{The dataset is publicly available at: \url{https://github.com/onwuegbuchulem/word_thumbnails_256_datasets/tree/main/Aptos}}. The generation pipeline replicates Windows \texttt{thumb.db} degradation using GPT-4 for content synthesis and ImageMagick to simulate downsampling and compression artifacts \cite{b20}.

The dataset generation pipeline was designed to rigorously replicate real-world forensic scenarios. As detailed in \cite{b20}, coherent text content is first generated using GPT-4 to mimic realistic document structure. This content is then formatted into Microsoft Word documents via the \texttt{python-docx} library and processed using the \texttt{wand} library (ImageMagick) to simulate the specific degradation characteristics of Windows \texttt{thumb.db} caches, including downsampling, blurring, and compression artifacts \cite{b20}. Each training instance consists of three components:
\begin{itemize}
    \item \textbf{Word Thumbnail}: The degraded input artifact. This is a low-resolution raster image that exhibits visual loss and compression characteristics of forensic thumbnail cache files.
    \item \textbf{Word Image}: The high-fidelity image. This serves as the spatial Ground Truth Image ($I_{\text{gt}}$), depicting how the document appears in its original, undegraded state.
    \item \textbf{Word Label}: The raw ASCII character string corresponding to the Word document content. This serves as the semantic Ground Truth Text ($T_{gt}$), providing the exact textual information contained within the image.
\end{itemize}

A primary challenge in training is managing the resolution disparity between the Word Thumbnail and the Word document. To address this issue, the preprocessing pipeline standardizes all images before they are entered into the network. Both the Word Thumbnail and the Word Image were resized to a uniform dimension of \textbf{$768 \times 594$ pixels} using bicubic interpolation. This preprocessing ensures a consistent mapping from degraded Word thumbnails to enhanced outputs, with all images converted to grayscale and normalized to the range \textbf{[-1, 1]} prior to training.

\section{Methodology}

\begin{figure*}[htbp]
  \centering
  \resizebox{0.95\textwidth}{!}{
  \begin{tikzpicture}[
      node distance=1.2cm and 1.5cm,
      font=\large\sffamily\bfseries,
      >={Latex[width=3mm,length=3mm]},
      netblock/.style={rectangle, draw=blue!60!black, fill=blue!5, thick, minimum width=3.2cm, minimum height=1.5cm, rounded corners, align=center, drop shadow},
      dashedblock/.style={rectangle, draw=blue!40!black, fill=blue!2, dashed, thick, minimum width=2.8cm, minimum height=1.2cm, rounded corners, align=center},
      bottle/.style={rectangle, draw=blue!80!black, fill=blue!10, thick, minimum width=3.0cm, minimum height=1.5cm, rounded corners, align=center, drop shadow},
      data/.style={rectangle, draw=gray, dashed, fill=white, minimum width=2.8cm, minimum height=1.2cm, align=center},
      loss/.style={rectangle, draw=red!60!black, fill=red!5, thick, minimum width=3.0cm, minimum height=1.4cm, rounded corners, align=center},
      mergedloss/.style={rectangle, draw=purple!60!black, fill=purple!5, thick, minimum width=3.0cm, minimum height=1.8cm, rounded corners, align=center, drop shadow},
      sum/.style={circle, draw=orange!60!black, fill=orange!5, thick, minimum size=1.6cm, align=center, font=\Huge\bfseries, drop shadow},
      update/.style={rectangle, draw=green!60!black, fill=green!5, rounded corners, minimum height=1.8cm, align=center, thick, drop shadow},
      preproc/.style={rectangle, draw=purple!60!black, fill=purple!5, thick, minimum width=2.8cm, minimum height=1.6cm, rounded corners, align=center, drop shadow},
      flow/.style={->, thick, draw=darkgray, line width=2.5pt, rounded corners=5pt},
      skip/.style={->, dashed, thick, draw=blue!50, line width=1.8pt},
      backprop/.style={->, dotted, ultra thick, draw=red!70, line width=3.0pt},
      logicflow/.style={->, dashed, thick, draw=orange!80!black, line width=2.0pt, rounded corners=5pt}
  ]

    \node[netblock] (enc1) {Encoder 1\\(64 ch)\\$512 \times 512$};
    \node[dashedblock, below=0.8cm of enc1] (enc_mid) {Hidden Layers\\(128, 256 ch)};
    \node[bottle, below right=0.8cm and 0.5cm of enc_mid] (bot) {Bottleneck\\(512 ch)\\$64 \times 64$};
    \node[dashedblock, above right=0.8cm and 0.5cm of bot] (dec_mid) {Hidden Layers\\(256, 128 ch)};
    \node[netblock, above=0.8cm of dec_mid] (dec1) {Decoder 1\\(64 ch)\\$512 \times 512$};

    \draw[flow] (enc1) -- (enc_mid);
    \draw[flow] (enc_mid.south) |- (bot.west);
    \draw[flow] (bot.east) -| (dec_mid.south);
    \draw[flow] (dec_mid) -- (dec1);

    \draw[skip] (enc1) -- node[midway, above, font=\normalsize\bfseries, text=blue!70] {Skip Connection} (dec1);
    \draw[skip] (enc_mid) -- node[midway, above, font=\small, text=blue!70] {Skip} (dec_mid);

    \node[data, left=2.0cm of enc1] (raw_input) {Word\\Thumbnail};
    \node[preproc, left=2.0cm of enc_mid] (resize) {Resize to\\$768 \times 594$};
    \node[data, below=1.2cm of resize] (input) {Resized\\Word Thumbnail};

    \draw[flow] (raw_input) -- (resize);
    \draw[flow] (resize) -- (input);
    \draw[flow] (input.east) -- ++(0.5,0) |- (enc1.west);

    \node[right=0.2cm of input, font=\footnotesize, align=left, text width=2.5cm, color=darkgray] 
    (normalization_text) {Grayscale \&\\Normalize $[-1, 1]$};

    \node[data, right=1.5cm of dec1] (restored) {Enhanced\\Word Thumbnail\\($I_{enhanced}$)};
    \draw[flow] (dec1) -- (restored);

    \node[loss, right=1.75cm of restored] (l1loss) {$L_{L1}$\\(Pixel Loss)};
    \node[data, above=0.5cm of l1loss] (gt_img) {Word \\Image ($I_{gt}$)};
    \draw[flow] (gt_img) -- (l1loss);

    \node[mergedloss, below right=1.75cm and 1.75cm of restored] (ocr_combined) {$L_{OCR}$\\(OCR Loss)};
    \node[data, below=0.5cm of ocr_combined] (gt_text) {Word Label ($T_{gt}$)};
    \draw[flow] (gt_text) -- (ocr_combined);

    \draw[flow] (restored.east) -- ++(1.5,0) |- (l1loss.west);
    \draw[flow] (restored.east) -- ++(1.0,0) |- (ocr_combined.west);


    \node[sum] (sigma) at ($(l1loss.east)!0.5!(ocr_combined.east)$) [xshift=1.5cm] {$\Sigma$};

    \draw[flow] (l1loss.east) -| node[pos=0.3, above] {$\lambda_{L1}$} (sigma.north);
    
    \draw[logicflow] (ocr_combined.east) -| node[pos=0.3, below right] {$\lambda_{OCR}$ (Selection)} (sigma.south);

    \node[update, right=1.0cm of sigma] (opt) {Optimizer\\(Update)};
    \draw[flow] (sigma) -- (opt);

    \draw[backprop] (opt.south) |- ++(0,-3.75cm) -| node[pos=0.25, below, font=\LARGE\bfseries, text=black] {Backpropagation Loop} ([xshift=-0.5cm]enc1.south west) -- (enc1.south west);

  \end{tikzpicture}
  }
  \caption{Architecture of the Text-Aware U-Net. The framework utilizes a hybrid evaluation strategy: the Visual Loss ($L_{L1}$) (top right) provides a differentiable gradient signal for backpropagation (solid path), whereas the Text-Aware Loss ($L_{OCR}$) (bottom right) serves as a non-differentiable validation metric to guide model selection (dashed path). This ensures the final model prioritizes semantic legibility without requiring end-to-end differentiability of the OCR engine.}
  \label{fig:u_shape_pipeline}
\end{figure*}
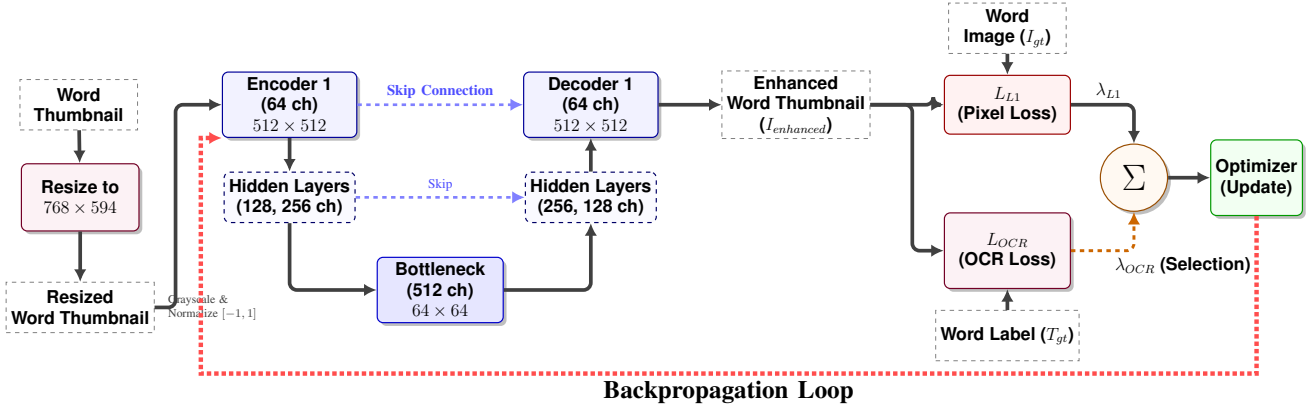

To address the challenge of restoring textual legibility in degraded forensic artifacts, we propose a deep learning framework based on a \textbf{U-Net architecture}. Standard super-resolution approaches typically rely on pixel-level convergence, often resulting in smooth but illegible text. To achieve text-aware capability, the proposed methodology employs a \textbf{Metric-Guided Optimization Strategy}. This approach decouples the training objective (gradient descent via pixel loss) from the evaluation objective (model selection via text readability), ensuring that the final model is optimized for forensic utility. The overall framework is illustrated in Fig.~\ref{fig:u_shape_pipeline}.


\subsection{Network Architecture}

The framework utilizes a symmetric U-Net \cite{b12} (Fig.~\ref{fig:u_shape_pipeline}) that down-samples inputs from $512 \times 512$ to a $64 \times 64$ bottleneck, expanding features from 64 to 512 channels to capture semantic patterns. Crucially, skip connections concatenate the encoder features directly to the decoder. This mechanism mitigates vanishing gradients and recovers fine-grained character strokes lost during downsampling, ensuring the high spatial fidelity and sharp text boundaries necessary for forensic analysis.

\subsection{Metric-Guided Optimization Strategy}

A significant challenge in training forensic enhancement models is the limitation of standard OCR engines (e.g., Tesseract) in a deep learning pipeline. These engines are non-differentiable, meaning that they process images using discrete logic (such as dictionary matching and character classification) rather than continuous mathematical functions. Consequently, standard backpropagation algorithms cannot calculate the gradients through the OCR engine to update the network weights.

This creates a conflict: we need to optimize text legibility, but we cannot train directly on text legibility. To solve this, we propose a dual-objective strategy in which Pixel Loss provides the mathematical gradients for learning, whereas OCR Loss acts as the definitive metric for model selection.

\subsubsection{Visual Loss Function ($\mathcal{L}_{L1}$)}
The training process is driven by a differentiable pixel-based L1 loss that measures the absolute difference between the enhanced output ($I_{\text{enhanced}}$) and the ground-truth image ($I_{\text{gt}}$):
\begin{equation}
    \mathcal{L}_{L1} = \frac{1}{N} \sum_{i=1}^{N} |I_{\text{enhanced}}^{(i)} - I_{\text{gt}}^{(i)}|
\end{equation}
This component provides the necessary gradient signal for backpropagation. By minimizing the pixel-wise distance, the network learns to reconstruct the global structure and contrast of a document, ensuring visual fidelity to the target.

\subsubsection{Text-Aware Model Selection ($\mathcal{L}_{OCR}$)}
To enforce semantic legibility without requiring a differentiable OCR architecture, we used the OCR metric as a strict validation criterion. During the validation phase, the enhanced Word Thumbnail is passed through the OCR engine to extract the recognized text $T_{\text{recog}}$. This is compared to the ground-truth label, $T_{\text{gt}}$, using the Normalized Levenshtein Distance (NLD):
\begin{equation}
    \mathcal{L}_{\text{OCR}} = \frac{\text{Levenshtein}(T_{\text{recog}}, T_{\text{gt}})}{\text{max}(|T_{\text{recog}}|, |T_{\text{gt}}|)}
\end{equation}

\begin{figure*}[htbp]
    \centering
    \begin{subfigure}[b]{0.333\textwidth}
        \centering
        \caption*{\textbf{Word Thumbnail}}
        \includegraphics[width=\textwidth, height=0.375\textheight]{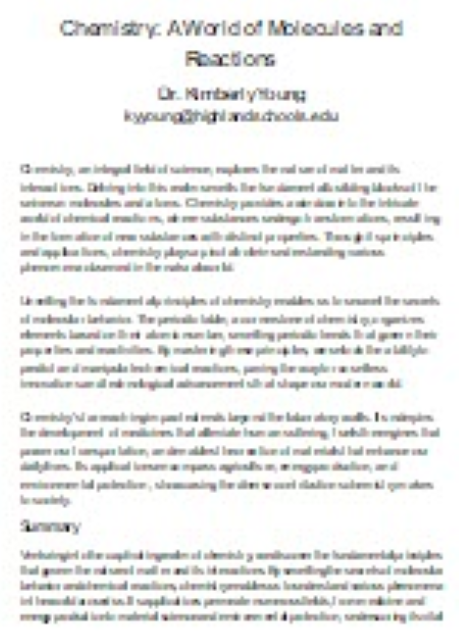}
        \label{fig:degraded2}
    \end{subfigure}%
    \begin{subfigure}[b]{0.333\textwidth}
        \centering
        \caption*{\textbf{Enhanced Word Thumbnail}}
        \includegraphics[width=\textwidth]{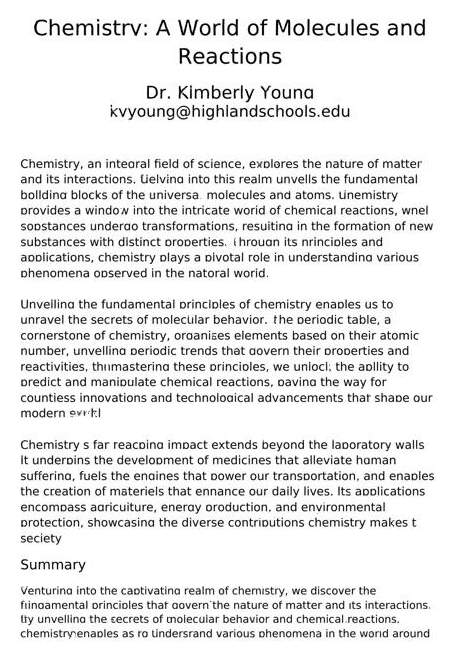}
        \label{fig:enhanced2}
    \end{subfigure}%
    \begin{subfigure}[b]{0.333\textwidth}
        \centering
        \caption*{\textbf{Word Image}}
        \includegraphics[width=\textwidth]{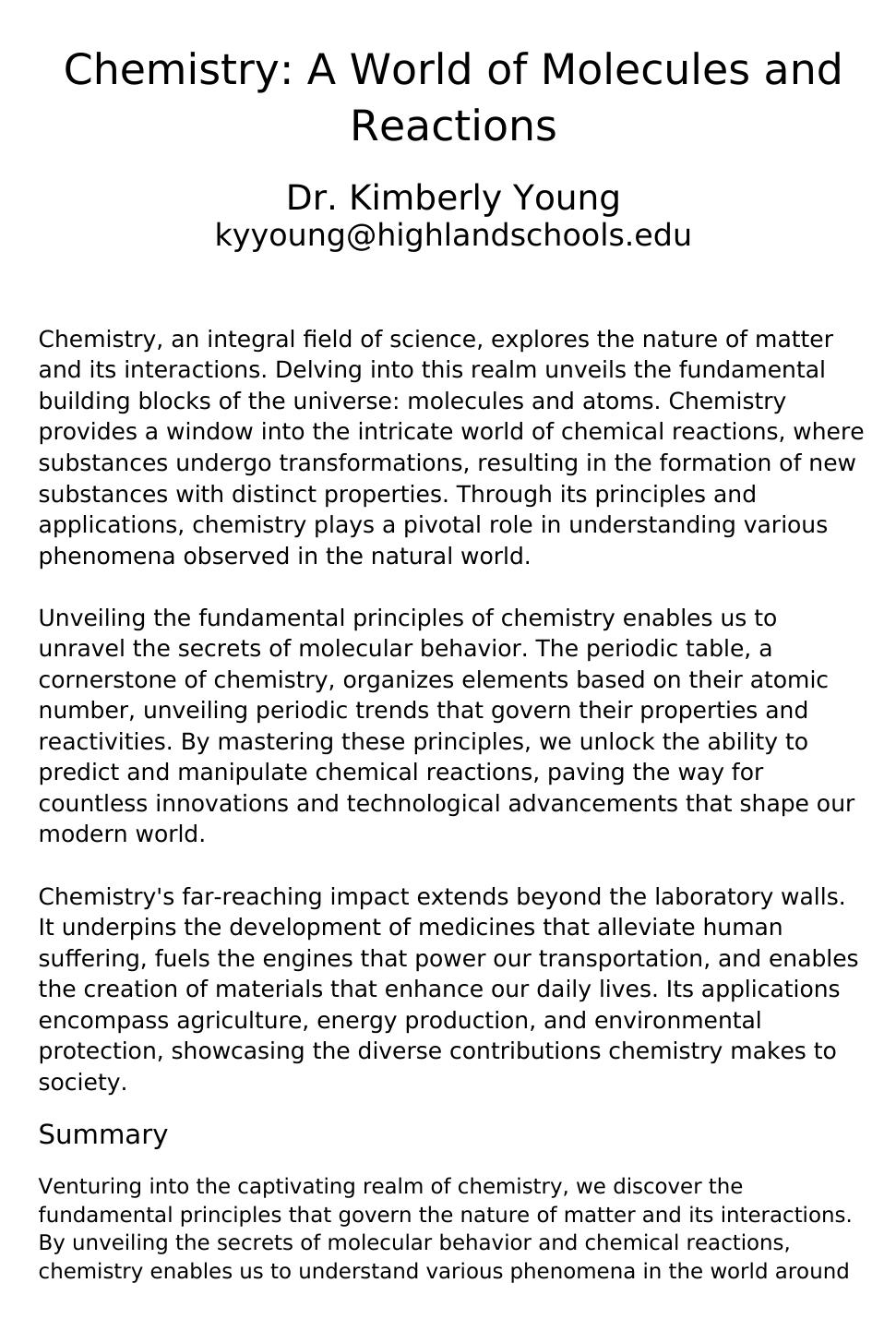}
        \label{fig:gt2}
    \end{subfigure}

    \caption{The figure presents a side-by-side evaluation of a \texttt{0326.jpg} thumbnail. \textbf{(Left)}~The Word Thumbnail ($768 \times 594$), exhibiting significant blur and compression artifacts. \textbf{(Center)}~The enhanced Word Thumbnail generated by the proposed Text-Aware U-Net, showing recovered character sharpness and legibility. \textbf{(Right)}~The word image for reference.}
    \label{fig:comparison_grid}
\end{figure*}

Instead of selecting the final model based solely on the lowest pixel error (which often prioritizes background smoothness over text sharpness), we employed a text-aware early stopping mechanism. The optimal model checkpoint is selected by minimizing the weighted validation score ($S_{\text{select}}$):
\begin{equation}
    S_{\text{select}} = \lambda_{L1}\mathcal{L}_{L1} + \lambda_{OCR}\mathcal{L}_{\text{OCR}}
\end{equation}
where $\lambda_{L1}$ and $\lambda_{OCR}$ control the relative importance of visual fidelity and textual legibility, respectively. This strategy ensures that the deployed model prioritizes the recovery of machine-readable text, effectively filtering out models that appear visually pleasant but fail to restore accurate character topology. A qualitative comparison illustrating the effect of this optimization strategy on degraded forensic thumbnails is shown in Fig.~\ref{fig:comparison_grid}.

\section{Empirical Study and Results}
\label{sec:empirical_study}

This section evaluates the effectiveness of the proposed text-aware restoration framework. We describe the experimental setup, evaluation metrics, and performance analysis using both quantitative measurements and qualitative inspection of the restored forensic artifacts.

\subsection{Experimental Setup}
The experiments were conducted using PyTorch on an NVIDIA A100 GPU. The proposed Text-Aware U-Net was trained for 550 epochs using the Adam optimizer with a fixed learning rate of $1 \times 10^{-4}$. The dataset consisted of 1,000 synthetic forensic thumbnails derived from \cite{b20}, partitioned into 70\% training and 30\% held-out test sets to evaluate generalization performance.

\textbf{Training Protocol:} Network parameters were optimized using pixel-wise L1 loss. To enforce text awareness without requiring a differentiable OCR engine, model checkpoints were selected using a weighted hybrid score,
$S_{\text{select}} = \lambda_{L1}\mathcal{L}_{L1} + \lambda_{OCR}\mathcal{L}_{\text{OCR}}$,
with $\lambda_{L1}=0.95$ and $\lambda_{OCR}=0.05$.

\subsection{Evaluation Metrics}
To provide a comprehensive assessment, we employ three distinct metrics that evaluate both the visual quality of the restored image and the machine-readability of its textual content.

\subsubsection{Visual Fidelity Metrics}

\textbf{Peak Signal-to-Noise Ratio (PSNR):} 
A widely used metric for measuring pixel-wise reconstruction quality. It is defined as:
\begin{equation}
\text{PSNR} = 20 \cdot \log_{10}\left(\frac{\text{MAX}_I}{\sqrt{\text{MSE}}}\right)
\end{equation}
where $\text{MAX}_I$ is the maximum possible pixel value (typically 255). The Mean Squared Error (MSE) quantifies the average squared deviation between the restored and target images. For an image of height $H$ and width $W$, MSE is given by:
\begin{equation}
\text{MSE} = \frac{1}{HW}
\sum_{i=1}^{H}\sum_{j=1}^{W}
\left( I_{\text{enhanced}}(i,j) - I_{\text{gt}}(i,j) \right)^2
\end{equation}
where $I_{\text{enhanced}}$ and $I_{\text{gt}}$ represent the pixel intensities of the enhanced thumbnail and ground-truth document, respectively.

\textit{\textbf{Example}:} For the evaluated sample \texttt{0326.jpg}, the reported PSNR is $30.9858$ dB. By inverting the PSNR equation ($10^{30.9858/20} \approx 35.40$), we derive an effective MSE of approximately $51.9$, representing the average pixel-wise reconstruction error for this sample.

\textbf{Structural Similarity Index Measure (SSIM):} 
Unlike PSNR, SSIM evaluates perceptual degradation based on luminance, contrast, and structure. The formula is defined as:
\begin{equation}
\label{eq:ssim}
\text{SSIM}(x, y) =
\frac{(2\mu_x\mu_y + C_1)(2\sigma_{xy} + C_2)}
{(\mu_x^2 + \mu_y^2 + C_1)
 (\sigma_x^2 + \sigma_y^2 + C_2)}
\end{equation}
where the components are calculated over local sliding windows: $\mu$ represents luminance, $\sigma$ represents contrast, and $\sigma_{xy}$ represents structural correlation. This metric is computed locally and averaged to produce a single global score.

\textit{\textbf{Example:}} For \texttt{0326.jpg}, the global SSIM is $0.991$, indicating that local text structures such as edges and strokes are well-preserved.

\begin{table}[t]
\captionsetup{labelsep=newline, justification=centering, font=sc}
\caption{\textsc{Quantitative Comparison of Visual Fidelity Metrics on the Thumbnail Dataset}}
\label{tab:benchmark_results}
\centering
\small

\begin{tabular*}{\columnwidth}{@{\extracolsep{\fill}}|l|c|c|c|c|}
\hline
\multirow{2}{*}{\textbf{Method}} &
\multicolumn{4}{c|}{\textbf{Evaluated Metrics}} \\ \cline{2-5}
& \multicolumn{2}{c|}{\textbf{In-Domain}} 
& \multicolumn{2}{c|}{\textbf{Cross-Domain}} \\ \cline{2-5}
& \textbf{PSNR} & \textbf{SSIM} & \textbf{PSNR} & \textbf{SSIM} \\
\hline
SRCNN \cite{b1}  & 14.25 & 0.695 & 10.45 & 0.494 \\ \hline
EDSR \cite{b9}   & 13.44 & 0.710 & 10.25 & 0.517 \\ \hline
ESRGAN \cite{b18}& 10.50 & 0.524 & 10.62 & 0.587 \\ \hline
TASuRe \cite{b2} & 15.62 & 0.351 & 15.62 & 0.351 \\ \hline
Conventional U-Net \cite{b24} & 16.42 & 0.347 & 13.47 & 0.521 \\ \hline
\textbf{Text-Aware U-Net} 
& \textbf{30.99} & \textbf{0.991} & \textbf{14.08} & \textbf{0.779} \\ \hline
\end{tabular*}

\vspace{3pt}
\begin{minipage}{\columnwidth}
\footnotesize
\textit{In-domain evaluation:} Models specifically optimized and validated on the target Aptos typeface within the forensic thumbnail dataset. \\
\textit{Cross-domain evaluation:} Testing on different fonts with the same degradation levels, without additional training.
\end{minipage}
\end{table}

\subsubsection{Textual Readability Metrics}

\textbf{Levenshtein Distance (LD):} 
The primary metric for evaluating machine-readability is the Levenshtein Distance between the OCR-recognized text ($T_{recog}$) and the ground-truth word label ($T_{gt}$). It measures the minimum number of single-character operations (insertions, deletions, substitutions) required to transform one string into the other:
\begin{equation}
LD(T_{recog}, T_{gt}) = \min_{ops} (I + D + S)
\end{equation}
Here, $\min_{ops}$ denotes the minimization over all possible sequences of edit operations, ensuring the metric reflects the optimal (shortest) transformation path rather than an arbitrary one. To ensure robust evaluation, the OCR output is normalized to lowercase and stripped of non-alphanumeric tokens before alignment.

\textit{\textbf{Example}:} For the enhanced output of thumbnail \texttt{0326.jpg}, the OCR engine required a raw edit distance of $L_e = 400$ to match the reference transcription.

\textbf{Normalized Levenshtein Distance (NLD):} 
To standardize the error rate independent of document length, we employ the NLD:
\begin{equation}
NLD = \frac{L_e}{C}
\end{equation}
where $L_e$ is the raw edit count and $C$ is the total character count of the ground truth. The NLD ranges from 0 (perfect match) to 1 (complete mismatch).

\textit{\textbf{Example}:} For thumbnail \texttt{0326.jpg}, with $L_e = 400$ edits and a document length of $C = 2177$ characters, the calculation yields $NLD = 400/2177 \approx 0.1837$. This signifies an approximate 18.4\% error rate, validating the semantic recoverability of the enhanced image.

\subsection{Results and Analysis}

The performance of the proposed text-aware algorithm was compared with that of the baseline models on a held-out test set comprising 30\% of the dataset.

\subsubsection{\textbf{RQ1: To what extent does the proposed text-aware model improve the visual quality (PSNR, SSIM) of severely degraded thumbnails compared to standard baseline models?}}


\begin{figure}[htbp]
\centering
\includegraphics[width=0.48\textwidth]{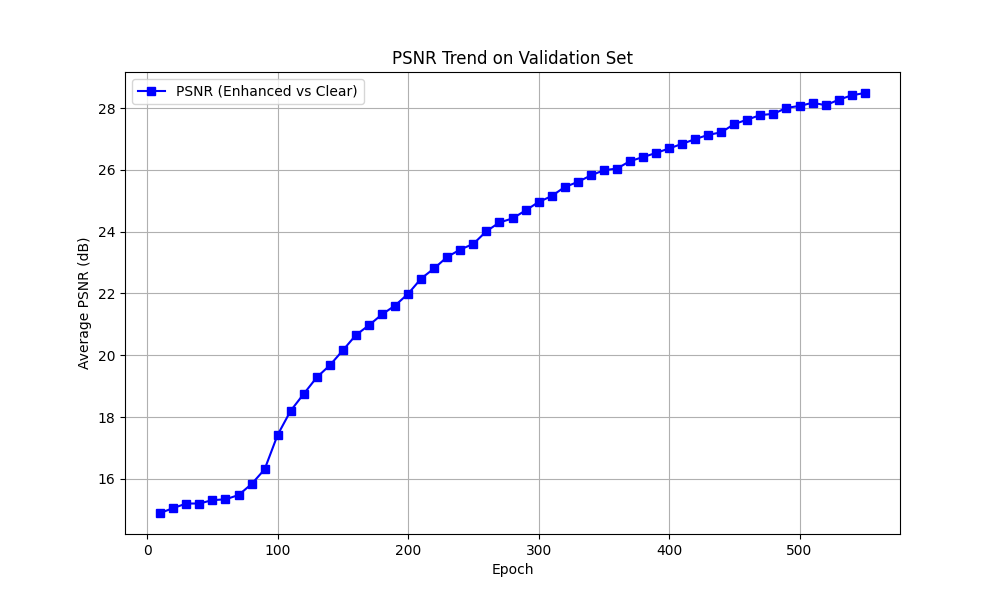}
\caption{PSNR trend on the validation set. The steady increase demonstrates consistent improvement in pixel-level fidelity.}
\label{fig:psnr_trend}
\end{figure}

\begin{figure}[H]
\centering
\includegraphics[width=0.48\textwidth]{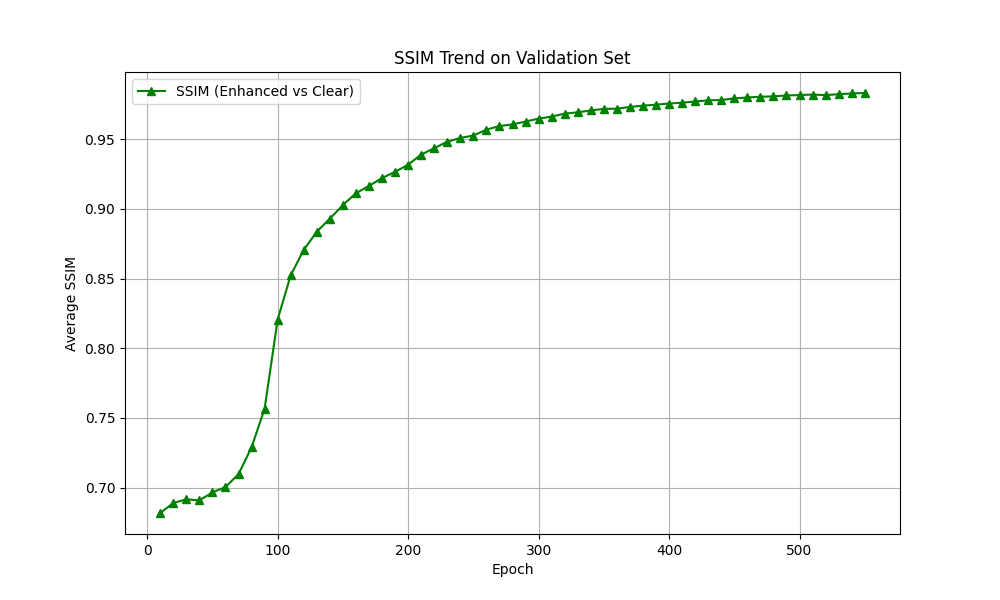}
\caption{SSIM trend on the validation set. The convergence toward 0.99 confirms the recovery of structural integrity.}
\label{fig:ssim_trend}
\end{figure}

To evaluate visual improvement, we analyzed the progression of visual fidelity metrics during training. As shown in Figure ~\ref{fig:psnr_trend}, the PSNR exhibits a monotonic upward trajectory, stabilizing at approximately 30.99 dB. This steady trend indicates that the model consistently minimizes pixel-wise reconstruction errors over successive epochs. 

The SSIM trend shown in Fig. ~\ref{fig:ssim_trend} shows a sharp increase, converging to a near-perfect value of 0.991. Unlike PSNR, which measures absolute pixel differences, SSIM captures the preservation of structural information that is critical for legible text reconstruction. This high SSIM score provides strong empirical evidence that the model does not merely smooth the image but actively recovers fine structural details, including the character edges and strokes.



\subsubsection{\textbf{Comparative Benchmarking}}

Table~\ref{tab:benchmark_results} compares the proposed method with representative baseline models. For in-domain evaluation, all baselines were retrained and evaluated on the same synthetic forensic thumbnail dataset derived from \cite{b20}. Under identical data conditions, conventional super-resolution 
GAN-based methods achieve moderate PSNR and SSIM values but struggle to preserve the character-level structure.


In contrast, the proposed Text-Aware U-Net attained substantially higher visual fidelity (PSNR = 30.99 dB, SSIM = 0.991). For cross-domain evaluation, pretrained baseline models were applied without fine-tuning, resulting in notable performance degradation. For the proposed method, cross-domain testing corresponds to evaluation on thumbnails generated with an unseen font family, while preserving 
same degradation pipeline. In this setting, performance degradation was limited (SSIM = 0.779), indicating robustness to font-level domain shifts.

\begin{figure}[htbp]
\centering
\includegraphics[width=0.48\textwidth]{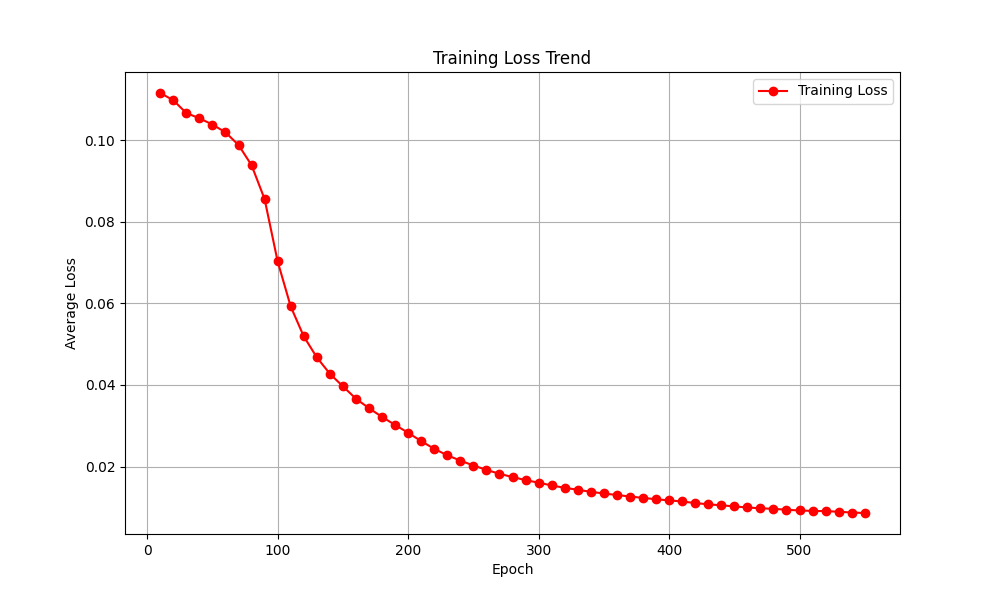} 
\caption{Training loss convergence over 550 epochs. The smooth decay confirms that the pixel-based optimization provides a stable gradient signal, while the OCR metric guides the selection of the optimal convergence point.}
\label{fig:loss_trend}
\end{figure}


\subsubsection{\textbf{RQ2: Does the integration of an OCR-based objective significantly improve text recoverability, as measured by Normalized Levenshtein Distance (NLD)?}}

To quantify the impact of the text-aware objective, the semantic performance was evaluated using the Normalized Levenshtein Distance (NLD). The network parameters were optimized via differentiable pixel-wise L1 minimization, as evidenced by the stable, oscillation-free convergence shown in Figure~\ref{fig:loss_trend}. The integration of the OCR-based objective as a \textbf{validation oracle} was decisive. This mechanism acts as a safeguard against the common failure mode of hallucinated or overly smoothed characters observed in purely pixel-driven super-resolution methods.


The effectiveness of this strategy was confirmed by quantitative results. The final selected model achieved an NLD of \textbf{0.184}, corresponding to a character recovery rate exceeding 81\%. Given the severe degradation of the input thumbnails, this result demonstrates that prioritizing text readability during model selection successfully bridges the gap between visual enhancement and machine readability, transforming previously illegible forensic artifacts into evidence that is suitable for automated analysis.

\section{Conclusion and Future Work}
\label{sec:conclusion}

This study addresses the critical forensic challenge of recovering legible text from severely degraded document thumbnails. By implementing a \textbf{U-Net} governed by a \textbf{Metric-Guided Optimization Strategy}, we overcome the limitations of conventional pixel-based super-resolution approaches. The proposed framework explicitly decouples the training objective, which emphasizes pixel fidelity, from the model selection criterion, which prioritizes the text readability. This design ensures that the final deployed model preserves the structural integrity required for a reliable forensic analysis.


The empirical impact of this approach was substantial. The proposed method achieves a Normalized Levenshtein Distance (NLD) of \textbf{0.184}, corresponding to a character recovery rate exceeding 81\% for Word Thumbnails. This performance significantly outperforms the standard baselines, demonstrating that prioritizing semantic recoverability during model selection is a decisive factor in transforming illegible thumbnail artifacts into actionable forensic evidence.

Future work will focus on integrating fully differentiable recognition modules, such as Convolutional Recurrent Neural Networks (CRNNs), to enable true end-to-end semantic optimization. In addition, we plan to validate the proposed framework on multilingual document datasets and investigate the use of Large Language Models (LLMs) for post-restoration error correction and semantic refinement.


\begin{thebibliography}{00}
\bibitem{b1}
C. Dong, C. C. Loy, K. He, and X. Tang, ``Image super-resolution using deep convolutional networks,'' IEEE Trans. Pattern Anal. Mach. Intell., vol. 38, no. 2, pp. 295--307, Feb. 2016.
\bibitem{b2}
E. Filonenko, A. Filonenko, and K.-H. Jo, ``TASuRe: Text aware super-resolution,'' in Proc. Int. Workshop Intell. Syst. (IWIS), 2022, pp. 1--5.
\bibitem{b3}
M. Hradiš, J. Kotera, P. Zemčík, and F. Šroubek, ``Convolutional neural networks for direct text deblurring,'' in Proc. Brit. Mach. Vis. Conf. (BMVC), 2015.
\bibitem{b4}
P. Isola, J.-Y. Zhu, T. Zhou, and A. A. Efros, ``Image-to-image translation with conditional adversarial networks,'' in Proc. IEEE Conf. Comput. Vis. Pattern Recognit. (CVPR), 2017, pp. 1125--1134.
\bibitem{b5}
K. Jiang, Q. Wang, Z. An, Z. Wang, C. Zhang, and C.-W. Lin, ``Mutual Retinex: Combining transformer and CNN for image enhancement,'' IEEE Trans. Emerg. Topics Comput. Intell., vol. 8, no. 3, pp. 2240--2252, Jun. 2024.
\bibitem{b6}
J. P. Khatiwala, D. K. N. Addai, and W. Xu, ``Evaluating the reliability of digital forensic evidence discovered by large language model: A case study,'' in Proc. IEEE COMPSAC, 2025, pp. 1067--1076.
\bibitem{b7}
C. Ledig \textit{et al.}, ``Photo-realistic single image super-resolution using a generative adversarial network,'' in Proc. IEEE CVPR, 2017, pp. 4681--4690.
\bibitem{b8}
V. I. Levenshtein, ``Binary codes capable of correcting deletions, insertions, and reversals,'' Soviet Physics Doklady, vol. 10, no. 8, pp. 707--710, 1966.
\bibitem{b9}
B. Lim, S. Son, H. Kim, S. Nah, and K. M. Lee, ``Enhanced deep residual networks for single image super-resolution,'' in Proc. IEEE CVPR Workshops, 2017.
\bibitem{b10}
S. Nah, T. H. Kim, and K. M. Lee, ``Deep multi-scale convolutional neural network for dynamic scene deblurring,'' in Proc. IEEE CVPR, 2017, pp. 3883--3891.
\bibitem{b11}
M. W. Nofan and O. Ata, ``Forensic image enhancement and forgery detection using CNNs,'' in Proc. ICOCSETI, 2025, pp. 548--553.
\bibitem{b12}
O. Ronneberger, P. Fischer, and T. Brox, ``U-Net: Convolutional networks for biomedical image segmentation,'' in MICCAI, 2015, pp. 234--241.
\bibitem{b13}
R. Smith, ``An overview of the Tesseract OCR engine,'' in Proc. ICDAR, 2007, pp. 629--633.
\bibitem{b14}
G. Sun and F. Zhang, ``CNN-based fast back projection imaging with noise-resistant capability,'' IEEE Access, vol. 8, pp. 117080--117086, 2020.
\bibitem{b15}
J. Wang, W. Wang, and T. Lu, ``Text-aware network for restoring degraded text images,'' in Proc. IEEE ICASSP, 2020, pp. 2513--2517.
\bibitem{b16}
W. Wang \textit{et al.}, ``Scene text image super-resolution in the wild,'' in Proc. ECCV, 2020, pp. 650--666.
\bibitem{b17}
X. Wang \textit{et al.}, ``TextSR: Content-aware text super-resolution guided by recognition,'' arXiv:1909.07113, 2019.
\bibitem{b18}
X. Wang \textit{et al.}, ``ESRGAN: Enhanced super-resolution generative adversarial networks,'' in Proc. ECCV Workshops, 2018.
\bibitem{b19}
Z. Wang \textit{et al.}, ``Image quality assessment: From error visibility to structural similarity,'' IEEE Trans. Image Process., vol. 13, no. 4, pp. 600--612, Apr. 2004.
\bibitem{b20}
E. Xu \textit{et al.}, ``Leveraging large language models for generating training datasets for text extraction from thumbnails,'' in Proc. IEEE COMPSAC, 2025, pp. 1089--1094.
\bibitem{b21}
S. W. Zamir \textit{et al.}, ``Multi-stage progressive image restoration,'' in Proc. IEEE/CVF CVPR, 2021, pp. 14821--14831.
\bibitem{b22}
Z. Zhong, J. Liu, and L. Jin, ``Towards robust image super-resolution via OCR-driven feedback,'' in Proc. ECCV, 2022, pp. 466--482.
\bibitem{b23}
T. Nagata, T. Goto, S. Motohashi, H. Chen, and R. Aoki, ``Improvement of robustness in blind image restoration method using failing detection process,'' J. Image Graph., vol. 7, no. 1, pp. 1--8, Mar. 2019.
\bibitem{b24}
J. Gurrola-Ramos, O. Dalmau, and T. E. Alarc\'on, ``A residual dense U-Net neural network for image denoising,'' IEEE Access, vol. 9, pp. 31742--31754, 2021.
\end{thebibliography}
\end{document}